\documentclass[twocolumn]{revtex4}
\usepackage{amsmath,dcolumn,feynmf,epsfig,CJK}
\begin{document}
\title{POSSIBLE GENERATION OF A $\mbox{\boldmath$\gamma$}$-RAY LASER BY
ELECTRONS \\WIGGLING IN A BACKGROUND LASER}
\author{QI-REN ZHANG }
\address{School of Physics, Peking University, Beijing 100871,
China}

\begin{abstract} The possibility of
$\gamma-$ray laser generation by the radiation of wiggling electrons
in an usual background laser is discussed.


{\bf PACS number(s)}: 42.55.Vc, 42.50.Ct, 12.20.-m
\end{abstract} \maketitle

\vskip0.3cm

\section{Introduction}
To extend the spectrum of lasers to the $\gamma$-ray range is a long
dream for both nuclear and laser physicists. It means fantastic
improvement of techniques, such as much more precise measurements of
space-time, holography for nanometer size or even smaller objects,
and so on. There are various proposals\cite{ba,zg,t,av,k1,k2,z3} for
its realization, including those of using nuclear decay and
condensed positronium decay and those of using $\gamma-$ray emitted
by wiggling charged particles. \cite{ba,zg} are review articles. We
preferred the way of using the radiations from charged particles
wiggling in a periodically bent crystal\cite{k1,k2,z3}. However,
this proposal assumes that the crystal is ideal. Contrarily, the
real crystals are not ideal. They have lacunae and impurities. This
point hurts the reliability of the proposal. Another problem is
that, the theory used in this proposal is not fundamental, which
also spoils its reliability. Moreover, to bend the crystal
periodically in small space-time is not easy. Here we propose a way
for the $\gamma$-ray laser generation, which keeps basic ideals of
this proposal, but without its shortcomings.

Physically a laser is a classical limit of electromagnetic wave,
simple and clean. It wiggles charged particles in a well-known way,
therefore is an ideal wiggler for further laser generation,
including those in the $\gamma$-ray range. The whole process may be
described by a fundamental theory, it is quantum electrodynamics,
the best theory in physics nowadays. We therefore reformed the
quantum electrodynamics\cite{z4} for this purpose, to see whether it
may offer a better way for the $\gamma$-ray laser generation or not.
The main change is to substitute the laser state for the usual
vacuum state, the latter was defined to be a state without any
photon and charged particle. The quantization of electromagnetic
field is now around a given laser state instead of the usual vacuum
state. Base wave functions for the electron field quantization are
changed to the solutions of Dirac equation for electron in the given
laser, instead of that in the usual vacuum. These new base wave
functions are periodically distorted, showing the quantum wiggling
of electrons in the laser. This is a new picture of quantum
electrodynamics, equivalent to its other pictures. As in the
original picture, the Hamiltonian of the system is also divided into
two parts. One is for the free motion, another is the interaction
between the free parts. But now, electrons are wiggling under the
action of the applied laser. Their freedom only means that they do
not emit or absorb photons. Photons now are quanta of the
fluctuation of the electromagnetic field relative to the given
laser. The laser, a given classical part of the electromagnetic
field is subtracted. This free Hamiltonian is exactly handled like
that in the original picture. Interaction Hamiltonian governs the
emission and absorption of single photon by an electron. It is
handled by perturbation. This is save because of the smallness of
the fine structure constant $\alpha$. We see that the proposal is on
a sound foundation. It is as sound as the usual form of the quantum
electrodynamics.

On this foundation we find that, in a head on collision between a
mono-energy electron beam and a laser, in a wide range of electron
energy and laser frequency, a beam of $\gamma-$ray appears on the
forward direction of the electron motion. It is almost completely
polarized, mono-directional and monochromatic. It is therefore
worthy to try the $\gamma-$ray laser generation in this way.

There has been a lot of works on the electron laser collisions, both
theoretical and experimental\cite{m,at,c,ma,f,db,s,da,gu}.
Theoretical works are based on the Compton formula and Klein-Nishina
formula for the Compton scattering\cite{c1,c2,kn,n}, which are
derived from the quantum electrodynamics in vacuum\cite{l,we}, but
without considering the stimulated emission of electrons. The
coherence of photon states in the laser\cite{v,a,b,z,g,le,zz} is
also overlooked. Experimental works found a way for generating
intense $\gamma-$ray by the Compton backscattering (CBS). However,
overlooking stimulated emissions, they overlooked the possibility of
making a $\gamma-$ray laser in this way. We hope these imperfections
may be remedied.

In section \ref{A2} we explain how a laser wiggles electrons and
makes them emit $\gamma-$quanta. The physics is emphasized.
Nevertheless, it is as quantitative as that given by the
systematically quantum electrodynamical derivation done in
\cite{z4}. Formulae for following calculations are given. In section
\ref{A3} we show numerical results obtained by these formulae,
explaining why we can generate $\gamma-$ray lasers in this way.
Section \ref{A4} is for the evolution of the $\gamma-$ray intensity,
because of the balance between the $\gamma-$quanta emission and
reabsorption. Possible intensity limit is discussed. Section
\ref{A5} is for proposals on further amplification of $\gamma-$ray
laser. Section \ref{A6} is for confirming the coherence of the
generated $\gamma-$ray laser. Section \ref{A7} is for conclusions.
\section{Quantum wiggling and photon emission
of electrons in a laser\label{A2}}

A laser is a classical limit of an intense electromagnetic wave and
is well described by a classical  4-potential. In the Coulomb gauge,
a circularly polarized laser is described by the vector potential
\begin{eqnarray}
\mbox{\boldmath $A$}(\mbox{\boldmath $x$})=A\{\mbox{\boldmath
$x$}_0\cos[k(z-t)] +\mbox{\boldmath $y$}_0 \sin[k(z-t)]\}\
.\label{16}
\end{eqnarray}
It is a plane wave circularly polarized in the $x-y$ plane and
propagating along the $z$ direction, with a wave vector
$\mbox{\boldmath $k$}=k\mbox{\boldmath $z$}_0$ and an amplitude $A$.
$\mbox{\boldmath $x$}_0$, $\mbox{\boldmath $y$}_0$ and
$\mbox{\boldmath $z$}_0$ are unit vectors along $x$, $y$ and $z$
directions respectively. The Dirac equation for an electron in this
laser is\begin{equation} {\rm i}\frac{\partial\psi}{\partial
t}=\{\mbox{\boldmath $\alpha$}\cdot(-{\rm
i}\mbox{\boldmath$\nabla$})+eA[\alpha_x\cos \phi+\alpha_y\sin
\phi]+\beta m\}\psi,\label{h}\end{equation} with $\phi\equiv
k(z-t)$. $e$ is the absolute value of the electron charge, and $m$
is the electron mass. Dirac matrices defined in Lurie's book\cite{l}
and the nature unit system ($c=\hbar=1$) are used. The operator on
the right hand side of this equation is time dependent. Fortunately,
a time dependent unitary transformation generated by the operator
${\rm e}^{-{\rm i}ktj_z}$ removes this time dependence\cite{z}, in
which
\begin{equation}j_z=-{\rm i}\frac{\partial}{\partial
\varphi}+\frac{\Sigma_z}{2}\end{equation} is the $z$-component of
the angular momentum of the electron. $\varphi$ is the azimuth angle
of the electron, defined to be the angle between the projection of
the electron radius vector on the $x-y$ plane and the $x$-axis.
$\Sigma_z$ is the $z$-component of the Pauli matrices. The
transformation
\begin{equation} \psi_r={\rm e}^{-{\rm i}ktj_z}\psi
\label{s0}\end{equation} is a rotation around the $z-$axis with
angular velocity $k$. The resulting picture is therefore called the
rotation picture, and denoted by the subscript $r$. Multiplying
${\rm e}^{-{\rm i}ktj_z}$ on two sides of equation (\ref{h}) from
left, we obtain the wave equation\begin{eqnarray} {\rm
i}\frac{\partial\psi_r}{\partial t}\!&=&\!\{\mbox{\boldmath
$\alpha$}\!\cdot\!(\!-{\rm i}\mbox{\boldmath$\nabla$})+
eA[\alpha_x\cos (kz)\!+\!\alpha_y\sin (kz)]+\beta m\nonumber
\\&+&kj_z\}\psi_r\label{gl}\end{eqnarray}in the rotation picture.
The operator on the right hand side of this equation is no longer
time dependent. We therefore have stationary solutions
$\psi_r(\mbox{\boldmath $x$},t)=U(\mbox{\boldmath $x$}){\rm
e}^{-{\rm i}\varepsilon t}$ for this equation. They satisfy the
eigen-equation
\begin{eqnarray}\!\!\! &&\{\mbox{\boldmath
$\alpha$}\!\cdot\!(\!-{\rm i}\mbox{\boldmath$\nabla$})+
eA[\alpha_x\cos (kz)\!+\!\alpha_y\sin (kz)]+\beta
m+kj_z\}U\nonumber\\&&=\varepsilon U,\label{g2}\end{eqnarray} in
which $U(\mbox{\boldmath $x$})$ is an eigenfunction and
$\varepsilon$ is the corresponding eigenvalue. This equation is
exactly solved\cite{z4,w}, with
\begin{equation}\varepsilon\equiv\varepsilon_n\equiv\varepsilon_{n,\sigma,\tau}(p_z,p_{_\perp})
=E+\frac{e^2A^2}{2(E-p_z)}
+(\frac{\sigma}{2}-n)k\label{i}\end{equation}and\begin{eqnarray}
\!\!\!\!\!\!\!\!\!&&U(\mbox{\boldmath$x$})\equiv
U_n(\mbox{\boldmath$x$})\equiv
U_{n,\sigma,\tau}(p_z,p_{_\perp};\mbox{\boldmath$x$})\nonumber\\\!\!\!\!\!\!\!\!\!&&=\frac{1}{2\pi}\exp\left\{{\rm
i}\left[p_z+\frac{e^2A^2}{2(E-p_z)}\right]z\right\}\nonumber
\\\!\!\!\!\!\!\!\!\!&&\times\!\!\left\{\!\!1\!\!-\!\!
\frac{eA}{2(p_z\!-\!E)}\!\left[\alpha_x\!\cos kz\!\!+ \!\alpha_y\sin
kz \!\!+\!{\rm i}(\Sigma_y\!\cos kz\!\!-\!\!\Sigma_x\!\sin
kz)\right]\!\right\}\nonumber\\\!\!\!\!\!\!\!\!\!&&\times\!\!\left[{\rm
i}^n\!{\rm J}_n\!(p_{_\perp}\rho'){\rm e}^{-{\rm i}n\varphi'}\!{\rm
P}\!_+\!\!+\!{\rm i}^{n\!-\!\sigma}\!{\rm
J}_{n-\sigma}\!(p_{_\perp}\rho'){\rm e}^{-{\rm
i}(n-\sigma)\varphi'}\!{\rm
P}\!_-\!\right]\!\!u_\sigma\!(0).\!\!\label{k}\end{eqnarray}
\mbox{\boldmath$p$}, $E=\tau\sqrt{p^2+m^2}$ with $\tau=\pm 1, \sigma
=\pm 1$ and $n\!=\mbox{integer}$ are parameters characterizing the
solution. Sometimes $n$ is chosen to be the representative of these
parameters. ${\rm J}_\nu(\xi)$ is a Bessel function of order $\nu$
in variable $\xi$. In the cylindric coordinates
$p_z,p_\perp,\varphi_p$ for the momentum $\mbox{\boldmath$p$}$, with
$p_x=p_\perp\cos\varphi_p$ and $p_y=p_\perp\sin\varphi_p$, the
bispinor factor in the plane wave solution of Dirac equation for a
free electron is\begin{eqnarray}u=u_+ +\,u_-{\rm e}^{{\rm
i}\sigma\varphi\!_p},\label{v6}\end{eqnarray}with\begin{eqnarray}
u_+\!\!=\!\!\sqrt{\frac{E+m}{2E}}\!\left[\!\!\begin{array}{c}1\\
\frac{p_z\sigma}{E+m}\end{array}\!\!\right]\!\!\chi\!_{_\sigma}
,\;\;u_-\!\!=\!\!\sqrt{\frac{E+m}{2E}}\!\left[\!\!\begin{array}{c}0\\
\frac{p_\perp}{E+m}\end{array}\!\!\right]\!\!\chi\!_{_{-\!\sigma}},\label{u}
\end{eqnarray}
in which $\chi_{_\sigma}$ is an eigenspinor of $\Sigma_z$ with
eigenvalue $\sigma$. For given $p_z$ and $p_\perp$, the bispinor
 (\ref{v6}) is characterized by $\sigma$ and $\varphi\!_p$, therefore may
be denoted by $u_\sigma(\varphi\!_p)$. The bispinor $u_\sigma(0)$ at
the end of eq.(\ref{k}) is defined in this way. The projection
operator ${\rm P}\!_\pm\!\!\equiv\!\!\frac{1\pm\sigma\Sigma_z}{2}$
gives ${\rm P}\!_\pm u_\sigma(0)\!=u_\pm$.   $(\rho',\varphi',z)$ in
wave function (\ref{k}) are coordinates of the electron position, in
which $(\rho',\varphi')$ are defined
by\begin{eqnarray}\left.\begin{array}{c}x'=x-\frac{eA}{k(p_z-E)}
\sin kz=\rho'\cos\varphi',\\y'=y+\frac{eA}{k(p_z-E)} \cos
kz=\rho'\sin\varphi' .\end{array}\right\}\end{eqnarray} These
equations constitute a $z$-dependent linear coordinate
transformation from Cartesian coordinates $(x,y,z)$ to $(x',y',z)$
in $x-y$ plane, followed by a transformation from rectangular
coordinates $(x',y')$ to polar coordinates $(\rho',\varphi')$ in
this plane. The factor $\frac{1}{2\pi}$ on the right hand side of
solution (\ref{k}) is a normalization constant. The set
$[U_{n,\sigma,\tau}(p_z,p_{_\perp};\mbox{\boldmath$x$})]$ of all
eigenfunctions is therefore orthonormal, so that
\begin{eqnarray}&&\int
U^\dag_{n,\sigma,\tau}(p_z,p_{_\perp};\mbox{\boldmath$x$})
U_{n',\sigma',\tau'}(p'_z,p'_{_\perp};\mbox{\boldmath$x$}){\rm d}^3x
\nonumber\\&&=\frac{1}{p_{_\perp}}\delta(p_{_\perp}-p'_{_\perp})
\delta(p_z-p'_z)\delta_{n,n'}\delta_{\sigma,\sigma'}\delta_{\tau,\tau'}.\label{g3}\end{eqnarray}
Since in this work we are interested only in the electron-laser
interaction, always have $\tau=1$, the eigenfunctions will be simply
denoted by $U_n(\mbox{\boldmath$x$})\equiv
U_{n,\sigma}(p_z,p_{_\perp};\mbox{\boldmath$x$}) \equiv
U_{n,\sigma,1}(p_z,p_{_\perp};\mbox{\boldmath$x$})$ in the
following.

At the limit $A\rightarrow 0$,
\begin{eqnarray}&&\!\!\!\!\!\!\!U_{n,\sigma}(p_z,p_{_\perp};\mbox{\boldmath$x$})\!\!\rightarrow\!\!
U^{(0)}_{n,\sigma}(p_z,p_{_\perp};\mbox{\boldmath$x$})\!\!\equiv\!\!\frac{{\rm
e}^{{\rm i}p_zz}}{2\pi}\!\!\left[{\rm i}^n{\rm
J}_n(p_{_\perp}\rho){\rm e}\!^{-{\rm i}n\varphi}{\rm
P}_+\right.\nonumber\\&&\!\!\!\!\!\!\!\!\left.+{\rm
i}^{n-\sigma}{\rm J}_{n-\sigma}(p_{_\perp}\rho){\rm e}^{-{\rm
i}(n-\sigma)\varphi}{\rm P}_-\right]\!\!u_\sigma(0),\end{eqnarray}
$\rho,\varphi,$ and $z$ are cylindric coordinates of the electron.
The right hand side is a solution of Dirac equation for a free
electron, and
\begin{eqnarray}\sum_{n=-\infty}^\infty\!\frac{{\rm e}^{{\rm i}n\varphi\!_p}}
{\sqrt{2\pi}}\,U^{(0)}_{n,\,\sigma}(p_z,p_{_\perp};\mbox{\boldmath$x$})=\frac{1}{\sqrt{(2\pi)^3}}\,{\rm
e}^{{\rm i}\mbox{\boldmath$p$}\cdot\,\mbox{\boldmath$x$}}u
\end{eqnarray}is a plane wave solution of Dirac equation for a free
electron of momentum
$\mbox{\boldmath$p$}=p_{_\perp}(\cos\varphi_p\mbox{\boldmath$x$}_0
+\sin\varphi_p\mbox{\boldmath$y$}_0)+p_z\mbox{\boldmath$z$}_0$, $u$
is defined by eq.(\ref{v6}). In the case of $p_{_\perp}=0$, we
have\begin{equation}U^{(0)}_{n,\,\sigma}(p_z,0;\mbox{\boldmath$x$})=\frac{\delta_{n,0}}{2\pi}{\rm
e}^{{\rm i}p_zz}u.\label{w1}\end{equation} It is nonzero only when
$n=0$. In this case, it is already a plane wave of electron with
momentum $\mbox{\boldmath$p$}=p_z\mbox{\boldmath$z$}_0$. Now we see
the influence of the laser on the electron motion. It is distortions
of the electron wave in $x-y$ planes, periodically along the $z$
direction, and the modulation of the periodicity along the $z$
direction. It is the wiggling of electrons. Since it appears in the
wave function instead in their trajectories, we call it quantum
wiggling.

The wiggling electron radiates because of the interaction
$e\mbox{\boldmath$\alpha$}\cdot\mbox{\boldmath$A$}'$ between the
electron and the electromagnetic field fluctuation around the laser.
$\mbox{\boldmath$A$}'$ is the difference between the total
electromagnetic vector potential and the vector potential (\ref{16})
of the laser. It is decomposed into various modes, each with a wave
vector $\mbox{\boldmath$k$}'$ and a polarization vector
$\mbox{\boldmath$e$}'$, perpendicular with each other in the Coulomb
gauge, and denoted by
$\iota'\equiv(\mbox{\boldmath$k$}',\mbox{\boldmath$e$}')$. It is
quantized. In the interaction picture, the wave function associated
with a photon of mode $\iota'$ is\cite{l}
\begin{equation}\mbox{\boldmath$A$}'_{\iota'} (\mbox{\boldmath$x$},t)=
\frac{\mbox{\boldmath$e$}'}{\sqrt{(2\pi)^32k'}}{\rm e}^{{\rm
i}(\mbox{\boldmath$k$}'\cdot\mbox{\boldmath$x$}-k't)}.\end{equation}
In the rotation and interaction picture, the interaction matrix
element for the electron transition from the state
$U_{n,\sigma}(p_z,0;\mbox{\boldmath$x$})$ to the state
$U_{n',\sigma'}(p'_z,p'_{_\perp};\mbox{\boldmath$x$})$ and emitting
a photon of wave vector $\mbox{\boldmath $k$}'$ and polarization
$\mbox{\boldmath $e$}'$ is\begin{eqnarray}&&\langle
p'_{_\perp},p'_z,\sigma',n';,\mbox{\boldmath$e$}',\mbox{\boldmath$k$}'
|H'_{(ri)}|n,\sigma,p_z,0\rangle\nonumber \\&&={\rm e}^{{\rm
i}[\varepsilon_{n'\!\!,\,\sigma'\!\!,1}(p'_z\!,\,p'_{_\perp})+k'\!-\varepsilon_{n\!,\sigma,1}(p_z\!,\,0)]t}
\!\!\int\!\!
U^\dag_{n',\sigma'}(p'_z,p'_{_\perp};\mbox{\boldmath$x$})\nonumber\\&&\times\frac{e\mbox{\boldmath
$\alpha$}\cdot\mbox{\boldmath $e$}'^*_r {\rm e}^{{-\rm
i}\mbox{\boldmath $k$}'_r\cdot\,\mbox{\boldmath
$x$}}}{\sqrt{(2\pi)^32k'}}U_{n,\sigma}(p_z,0;\mbox{\boldmath$x$}){\rm
d}^3x.\label{v}\end{eqnarray}In which\begin{eqnarray}\mbox{\boldmath
$k$}'_r&=&[k'_x\cos(kt)-k'_y\sin(kt)]\mbox{\boldmath
$x$}_0\nonumber\\&+& [k'_x\sin(kt)+k'_y\cos(kt)]\mbox{\boldmath
$y$}_0+k'_z\mbox{\boldmath
$z$}_0\label{t1}\end{eqnarray}and\begin{eqnarray} \mbox{\boldmath
$e$}'_r&=&[e'_x\cos(kt)-e'_y\sin(kt)]\mbox{\boldmath
$x$}_0\nonumber\\&+& [e'_x\sin(kt)+e'_y\cos(kt)]\mbox{\boldmath
$y$}_0+e'_z\mbox{\boldmath $z$}_0\label{t2}\end{eqnarray}are
$\mbox{\boldmath $k$}'$ and $\mbox{\boldmath $e$}'$ respectively in
the rotation picture. They rotate around the $z$ axis with an
angular velocity $k$. The rotation transformation makes the
unperturbed Hamiltonian of an electron in the background laser time
independent, and makes the perturbation of the electromagnetic field
fluctuation on the electron periodically time dependent. These two
changes work together results in the equivalence of the rotation
picture with the original one, and makes the problem able to be
handled by the usual time dependent perturbation.

Take
$\mbox{\boldmath$e$}'_1=\cos\theta(\cos\varphi\!_{_{k'}}\mbox{\boldmath$x$}_0
+\sin\varphi\!_{_{k'}}\mbox{\boldmath$y$}_0)-\sin\theta\mbox{\boldmath$z$}_0$
and $\mbox{\boldmath$e$}'_2=
-\sin\varphi\!_{_{k'}}\mbox{\boldmath$x$}_0+\cos\varphi\!_{_{k'}}\mbox{\boldmath$y$}_0$
to be a pair of orthonormal polarization vectors orthogonal to the
wave vector $\mbox{\boldmath$k$}'$, in which $\theta$ is the angle
between $\mbox{\boldmath$k$}'$ and the $z$ axis, $\varphi_{_{k'}}$
is the angle between the projection of $\mbox{\boldmath$k$}'$ on the
$x-y$ plane and the $x$ axis. The integral in the eq.(\ref{v}) has
been analytically worked out\cite{z4}. It
gives\begin{eqnarray}&&\langle
p'_{_\perp},p'_z,\sigma',n';\mbox{\boldmath$e$}'_i,\mbox{\boldmath$k$}'
|H'_{(ri)}|n,\sigma,p_z,0\rangle
=\frac{e\delta(p'_{_\perp}-k'_{_\perp})}{\sqrt{(2\pi)^32k'}
p'_{_\perp}}\nonumber\\&&\times\sqrt{\frac{1}{4EE'(E+m)(E'+m)}}\sum_{{\cal
N}=-\infty}^\infty\sum_{\nu=0,\pm1}{\rm J}_{{\cal N}-\nu}(p'_\perp
R') \nonumber\\&&\times\left[\delta_{\sigma,\sigma'}F^{(\nu)}_i{\rm
e}^{-{\rm i}{\cal
N}\varphi\!_{_{k'}}}+\delta_{\sigma,-\sigma'}G^{(\nu)}_i{\rm
e}^{{\rm i}(\sigma-{\cal
N})\varphi\!_{_{k'}}}\right]\nonumber\\&&\times\delta
[p'_z-p_z+\frac{eA}{2}(R'-R)k+k'_z-{\cal N}
k]\nonumber\\&&\times{\rm e}^{{\rm i}n'(\varphi\!_{_{k'}}+\pi)}{\rm
e}^{{\rm i} [E'-E+\frac{eA}{2}(R'-R)k+k'-{\cal N}
k]t},\label{g4}\end{eqnarray}with $R\equiv\frac{eA}{k(E-p_z)}$,
$R'\equiv\frac{eA}{k(E'-p'\!_z)}$,\begin{eqnarray}\left.\begin{array}{l}F^{(0)}_1\equiv-\cos\theta
p'_\perp(E+m)-\sin\theta[(E'+m)p_z\\+\!(E+m)p'_z\!\!
+\!\frac{1}{2}k^2RR'(p_z\!-\!E\!-\!m)(p'_z\!-\!E'\!-\!m)],\\F^{(\sigma)}_1\equiv\frac{k}{2}\{\cos\theta
R (p_z-\!E-\!m)(p'_z-\!E'-m)\\+\!\sin\theta
p'_\perp[(R\!+\!R')(E+m)\!-\!(R\!-\!R')p_z\},\\
F^{(-\sigma)}_1\equiv\frac{k}{2}\cos\theta
R'(p_z-E-m)(p'_z-E'-m),\\G^{(0)}_1\equiv\sigma\{\cos\theta[p_z(E'+m)-p'_z(E+m)]
\\+\sin\theta p'_\perp[\frac{k^2RR'}{2}(p_z-E-m)+E+m]\},\\
G^{(\sigma)}_1\equiv -\frac{\sigma k}{2}\{\cos\theta
Rp'_\perp(p_z-E-m)\\+\sin\theta[R(p_z-E-m)(p'_z+E'+m)\\-R'(p_z+E+m)(p'_z-E'-m)]\}
,\\
G^{(-\sigma)}_1\equiv-\frac{\sigma k}{2}\cos\theta
R'p'_\perp(p_z-E-m),\end{array}\!\!\!\!\right\}\label{g7}\end{eqnarray}and
\begin{eqnarray}\left.\begin{array}{l}F^{(0)}_2\equiv -{\rm i}\sigma p'_\perp(E+m),
\\F^{(\sigma)}_2\equiv -{\rm i}\sigma\frac{kR}{2}(p_z-E-m)(p'_z-E'-m),\\
F^{(-\sigma)}_2\equiv{\rm
i}\sigma\frac{kR'}{2}(p_z-E-m)(p'_z-E'-m),\\G^{(0)}_2\equiv{\rm i}[p_z(E'+m)-p'_z(E+m)],\\
G^{(\sigma)}_2\equiv{\rm i}\frac{kR}{2}p'_\perp(p_z-E-m)
,\\
G^{(-\sigma)}_2\equiv -{\rm
i}\frac{kR'}{2}p'_\perp(p_z-E-m).\end{array}\;\;\;\;\;\;\right\}\label{g8}\end{eqnarray}
${\cal N}$ is an integer, its appearance and the summation over it
in eq. (\ref{g4}) come from the integration.

Suppose a free electron of momentum
$\mbox{\boldmath$p$}=p_z\mbox{\boldmath$z$}_0$ and spin $\sigma/2$
comes from remote past and meets a laser on the way. It evolves
according to Gell-Mann Low theorem\cite{gl} into the state
$U_{n,\,\sigma}(p_z,0;\mbox{\boldmath$x$})$. The above analysis
shows, this state may transit to a superposition
$\sum_{n'=-\infty}^\infty \frac{{\rm e}^{{\rm i}n'\varphi\!_{_p}}}
{\sqrt{2\pi}}
U_{n',\,\sigma'}(p'_z,p'_{_\perp};\mbox{\boldmath$x$})$ of electron
states in the laser due to the electromagnetic interaction
$H'_{(ri)}$, and emit a photon of momentum $\mbox{\boldmath$k$}'$.
The superposed electron state evolves once again in the laser, into
the state $\frac{1}{\sqrt{(2\pi)^3}}\,{\rm e}^{{\rm
i}\mbox{\boldmath$p$}'\cdot\,\mbox{\boldmath$x$}}u_{\sigma'}$ of a
free electron when goes to the remote future. In the initial state,
besides an electron in the state
$U_{n,\,\sigma}(p_z,0;\mbox{\boldmath$x$})$, there may be $N$
photons in the mode $(\mbox{\boldmath$k$}',\mbox{\boldmath$e$}'_i)$.
It is denoted by
$|N,\mbox{\boldmath$k$}',\mbox{\boldmath$e$}'_i;\mbox{\boldmath$p$},\sigma\rangle$,
with $\mbox{\boldmath$p$}=p_z\mbox{\boldmath$z$}_0$. In the final
state, the electron state has transited to $\sum_{n'=-\infty}^\infty
\frac{{\rm e}^{{\rm i}n'\varphi\!_{_p}}} {\sqrt{2\pi}}
U_{n',\,\sigma'}(p'_z,p'_{_\perp};\mbox{\boldmath$x$})$, and the
number of the photon in mode
$(\mbox{\boldmath$k$}',\mbox{\boldmath$e$}'_i)$ becomes $N+1$. It is
denoted by
$|N+1,\mbox{\boldmath$k$}',\mbox{\boldmath$e$}'_i;\mbox{\boldmath$p$}',\sigma'\rangle$.
The interaction matrix element is now\begin{eqnarray}&&\langle
\sigma',\mbox{\boldmath$p$}';\mbox{\boldmath$e$}'_i,\mbox{\boldmath$k'$},N+1
|H'_{(ri)}|N,\mbox{\boldmath$k$}',\mbox{\boldmath$e$}'_i;\mbox{\boldmath$p$},\sigma\rangle
=\frac{e}{2\pi\sqrt{2k'}
}\nonumber\\&&\times\sqrt{\frac{N+1}{4EE'(E+m)(E'+m)}}\sum_{{\cal
N}=-\infty}^\infty\sum_{\nu=0,\pm1}{\rm J}_{{\cal N}-\nu}(p'_\perp
R') \nonumber\\&&\times\left[\delta_{\sigma,\sigma'}F^{(\nu)}_i{\rm
e}^{-{\rm i}{\cal
N}\varphi\!_{_{k'}}}+\delta_{\sigma,-\sigma'}G^{(\nu)}_i{\rm
e}^{{\rm i}(\sigma-{\cal
N})\varphi\!_{_{k'}}}\right]\nonumber\\&&\times\delta
[\mbox{\boldmath$p'+k'-p$}-{\cal N}
\mbox{\boldmath$k$}+\frac{eA}{2}(R'-R)\mbox{\boldmath$k$}]\nonumber\\&&\times{\rm
e}^{{\rm i} [E'-E+\frac{eA}{2}(R'-R)k+k'-{\cal N}
k]t},\label{gs1}\end{eqnarray} in which the factor $\sqrt{N+1}$
makes the stimulated emission, \begin{eqnarray}&&\delta
[\mbox{\boldmath$p'+k'-p$}-{\cal N}
\mbox{\boldmath$k$}+\frac{eA}{2}(R'-R)\mbox{\boldmath$k$}]\nonumber\\&&=\frac{\delta(p'_{_\perp}-k'_{_\perp})}{p'_{_\perp}}
\delta(\varphi\!_{_{k'}}+\pi-\varphi\!_{_{p'}})\nonumber\\&&\times\delta
[p'_z-p_z+\frac{eA}{2}(R'-R)k+k'_z-{\cal N} k]
\end{eqnarray}is a 3-dimensional $\delta$-function.

In the first order perturbation, the transition amplitude of the
process is\begin{eqnarray}&&\langle
\sigma',\mbox{\boldmath$p$}';\mbox{\boldmath$e$}'_i,\mbox{\boldmath$k'$},N\!\!+\!\!1
|T|N,\mbox{\boldmath$k$}',\mbox{\boldmath$e$}'_i;\mbox{\boldmath$p$},\sigma\rangle
\nonumber\\&=&\!\!-{\rm i}\!\!\int_{-\infty}^\infty\!\!\!\!\langle
\sigma',\mbox{\boldmath$p$}';\mbox{\boldmath$e$}'_i,\mbox{\boldmath$k'$},N\!\!+\!\!1
|H'_{(ri)}|N,\mbox{\boldmath$k$}',\mbox{\boldmath$e$}'_i;\mbox{\boldmath$p$},\sigma\rangle
{\rm d}t\nonumber\\&=&\!\!-{\rm i}\frac{e}{\sqrt{2k'} }
\sqrt{\frac{N+1}{4EE'(E+m)(E'+m)}}\nonumber\\&&\times\sum_{{\cal
N}=-\infty}^\infty\!\sum_{\nu=0,\pm1}\!\!{\rm J}_{{\cal
N}-\nu}(p'_\perp R')\nonumber\\&&
\times\left[\delta_{\sigma,\sigma'}F^{(\nu)}_i{\rm e}^{-{\rm i}{\cal
N}\varphi\!_{_{k'}}}+\delta_{\sigma,-\sigma'}G^{(\nu)}_i{\rm
e}^{{\rm i}(\sigma-{\cal
N})\varphi\!_{_{k'}}}\!\!\right]\nonumber\\&&\times\delta
[\mbox{\boldmath$p'+k'-p$}-{\cal N}
\mbox{\boldmath$k$}+\frac{eA}{2}(R'-R)\mbox{\boldmath$k$}]\nonumber\\&&\times\delta[E'+k'-E-{\cal
N} k+\frac{eA}{2}(R'-R)k].\end{eqnarray}$\delta-$functions give
selection rules for non-zero transition probability. They are
\begin{equation}\mbox{\boldmath$p'+k'-p$}-{\cal N}
\mbox{\boldmath$k$}+\frac{eA}{2}(R'-R)\mbox{\boldmath$k$}=0\label{w5}\end{equation}and
\begin{equation}E'+k'-E-{\cal
N} k+\frac{eA}{2}(R'-R)k=0.\label{w}\end{equation}In the limit
$A=0$, they like the usual momentum and energy conservation of the
process, if ${\cal N}$ is interpreted to be the number of photons in
the laser which participate the collision with the electron.
However, from the derivation\cite{z4} we see this interpretation is
questionable. The appearance of ${\cal N}$ is due to the periodic
distortion of the electron wave, it is the quantum wiggling of the
electron in the laser. The exact meaning of ${\cal N}$ is therefore
the order of a harmonic component in the distorted electron wave.
The $A-$dependent terms in them show the coherence effect of the
laser. Using them, we obtain
\begin{eqnarray}k'\!\!=\!\!\frac{{\cal N} k(E-p_z)}{E+{\cal N} k+\frac{eA}{2}Rk-(p_z
+{\cal N} k+\frac{eA}{2}Rk)\cos\theta}\,
.\label{g5}\end{eqnarray}For given collinear incident electrons and
background laser, this formula gives the direction dependence of the
emitted photon energy.  In the case of ${\cal N}=1$ and $A=0$, it
reduces to the Compton formula
\begin{eqnarray}k_0'\!\!=\!\!\frac{k(E-p_z)}{E+k-(p_z +
k)\cos\theta}\label{g55}\end{eqnarray}of the  corresponding  Compton
scattering\cite{c1,c2}. $k_0'$ is the energy of the outgoing photon
when $A=0$.

The transition probability per-unit time in unit volume and unit
solid angle of $\mbox{\boldmath$k$}'$
is\cite{z4}\begin{eqnarray}&&\!\!\!\!\!\!\frac{\partial^5
P}{\partial^3x\partial t\partial\Omega_{k'}}=\frac{\alpha
k'}{(2\pi)^3}\frac{N+1}{4EE'(E+m)(E'+m)}\nonumber\\&&\!\!\!\!\!\!\times\frac{E'k'}{{\cal
N} k(E-p_z)} \!\!\sum_{{\cal N}=-\infty}^\infty\!\!
\left|\sum_{\nu=0,\pm1}\!\!\!\!{\rm J}_{{\cal N}-\nu}(p'_\perp
R')\!\! \left[\delta_{\sigma,\sigma'}F^{(\nu)}_i{\rm e}^{-{\rm
i}{\cal
N}\varphi\!_{_{k'}}}\right.\right.\nonumber\\&&\!\!\!\!\!\!\left.\left.+\delta_{\sigma,-\sigma'}G^{(\nu)}_i{\rm
e}^{{\rm i}(\sigma-{\cal
N})\varphi\!_{_{k'}}}\!\!\right]\right|^2.\label{g6}\end{eqnarray}
In our unit system, the compton wavelength of an electron is
$m^{-1}$. In the unit of $m^{-2}$, the differential cross section of
a piece of laser of volume $m^{-3}$, in which a photon of momentum
$\mbox{\boldmath$k$}'$ and polarization $\mbox{\boldmath$e$}'_i$ is
emitted in a unit solid angle by an incident electron, is\cite{z4}
\begin{eqnarray} &&\!\!\!\!\!\!\!\!\!\!\!\!\!\!\frac{{\rm d} \Sigma}{{\rm
d}\Omega_{k'}}(\mbox{\boldmath$e$}'_i)\!=\!\frac{\alpha
k'^2(N+1)}{8\pi m {\cal N} k |p_z|(E-p_z)(E+m)(E'+m)}
\nonumber\\&&\!\!\!\!\!\!\!\!\!\!\!\!\!\!\times\!\!\!\!\!\!\sum_{{\cal
N}=-\infty}^\infty\! \!\left|\!\sum_{\nu=0,\pm1}\!\!\!\!\!{\rm
J}_{{\cal N}-\nu}(p'_\perp R')\!\!
\left[\delta_{\sigma,\sigma'}F^{(\nu)}_i\!\!\!
+\!\delta_{\sigma,-\sigma'}G^{(\nu)}_i\!{\rm e}^{{\rm
i}\sigma\varphi\!_{_{k'}}}\!\!\right]\!\right|^2\!\!\!\!.\label{w2}\end{eqnarray}

For an arbitrarily given polarization
$\mbox{\boldmath$e$}'=\sum_{i=1,2}c_i\mbox{\boldmath$e$}'_i$ of the
created photon, with $|c_1|^2+|c_2|^2=1$, the cross section
(\ref{w2}) becomes
\begin{eqnarray} &&\!\!\!\!\frac{{\rm d} \Sigma}{{\rm
d}\Omega_{k'}}(\mbox{\boldmath$e$}')=\frac{\alpha k'^2(N+1)}{8\pi m
{\cal N} k |p_z|(E-p_z)(E+m)(E'+m)}
\nonumber\\&&\!\!\!\!\times\!\!\!\sum_{{\cal
N}=-\infty}^\infty\left|
\left[\delta_{\sigma,\sigma'}\mbox{\boldmath$e$}'^*\cdot\mbox{\boldmath${\cal
F}$}\!\!\!_{_{\cal N}}
+\delta_{\sigma,-\sigma'}\mbox{\boldmath$e$}'^*\cdot\mbox{\boldmath${\cal
G}$}\!_{_{\cal N}}\right]\right|^2,\label{wa}\end{eqnarray} in which
\begin{eqnarray}\left.\begin{array}{l}\mbox{\boldmath${\cal
F}$}\!\!_{_{\cal N}}=\sum_{i=1,2}\sum_{\nu =0,\pm 1}F_i^{(\nu)}{\rm
J}_{{\cal N}-\nu}(p'_\perp R')\mbox{\boldmath$e$}'_i\, ,
\\\mbox{\boldmath${\cal G}$}\!_{_{\cal N}}=\sum_{i=1,2}\sum_{\nu =0,\pm
1}G_i^{(\nu)}{\rm e}^{{\rm i}\sigma\varphi\!_{_{k'}}}{\rm J}_{{\cal
N}-\nu}(p'_\perp R')\mbox{\boldmath$e$}'_i\, .\end{array}\right\}
\end{eqnarray}
For definite $\sigma$, $\sigma'$ and ${\cal N}$, the polarization of
the outgoing photon is also definite. It is
$\mbox{\boldmath$e$}'_f\equiv\mbox{\boldmath${\cal F}$}_{\cal
N}/{\cal F}_{\cal N}$ for $\sigma'=\sigma$ or
$\mbox{\boldmath$e$}'_g\equiv\mbox{\boldmath${\cal G}$}_{\cal
N}/{\cal G}_{\cal N}$ for $\sigma'=-\sigma$, with ${\cal F}_{\cal
N}\equiv\sqrt{\mbox{\boldmath${\cal F}$}_{\cal
N}^*\cdot\mbox{\boldmath${\cal F}$}_{\cal N}}$ and ${\cal G}_{\cal
N}\equiv\sqrt{\mbox{\boldmath${\cal G}$}_{\cal
N}^*\cdot\mbox{\boldmath${\cal G}$}_{\cal N}}$.

\section{$\gamma-$emission and $\gamma-$ray laser generation from wiggling
 electrons in an usual laser\label{A3}}

 Consider the head on collision between an electron beam and a
 circularly polarized laser. To be definite, we take the Titanium Sapphire Laser
 of wave length 660nm to 1180nm tunable. Take the middle,
 we set the wave length of the laser to be 785nm. This kind of facility may be found
 in many laboratories over the world. By compression, the pulse intensity of
 this kind of laser may be as high as $2\times10^{26}\mbox{W/m}^2$\cite{oe}. However, the
 theory derived above is valid only when the applied laser may be considered as a plane wave.
 The linear dimensions of the laser pulse have to be much larger
 than the wave length. It means the laser should not be over
 compressed.  We would take a moderate intensity. It is
 $10^{19}\mbox{W/m}^2$. The energy of the incident electrons is
 limited in the range $10^2\;\mbox{to}\;10^3$MeV. It may be found
 in many laboratories too.

\begin{figure}\includegraphics[width=8cm]{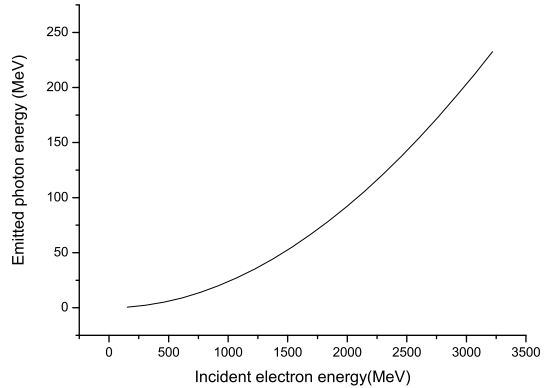}
\caption{Relation between the energy of the emitted photon on the
forward direction of the incident electron motion and the energy of
the incident electron, in the case of $\lambda=$785nm and
$I=10^{19}\mbox{W/m}^2$ for the laser.}\label{fig1}
\end{figure}

Now, a circularly polarized plane wave laser of wave length 785nm
propagates along the $z$-direction. A beam of electrons, each with
energy $E$, moves towards $-z$ direction. Electrons wiggle in the
laser and emit photons. Equations (\ref{g6}) and (\ref{w2}) show
that the emission probability decreases rapidly with the increase of
the integer ${\cal N}$, because of the contribution of Bessel
functions. We therefore confine ourselves in the case of ${\cal
N}=1$. Using (\ref{g5}) we may calculate the energy of the emitted
photon. To do this we need the coherence amplitude $A$ of the
applied laser. If the radiation is fully coherent, we have
\begin{equation}\frac{eA}{mc}=\sqrt{\frac{\alpha \lambda_c\lambda^2 I}
{\pi mc^3}}\, ,\label{g9}\end{equation} in which $\lambda$ is the
wavelength and $I$ is the intensity of the laser.
$\lambda_c=\hbar/mc=0.386$pm is the Compton wavelength of the
electron. Fixing $I=10^{19}\mbox{W/m}^2$ and $\lambda=785$nm, we get
$A=1.5\times10^{-2}mc/e$. Figure \ref{fig1} shows the calculated
relation between the energy of the emitted photon on the forward
direction of the incident electron motion and the energy of the
incident electron. We see the energy of the emitted photons are in
the $\gamma-$ray range.

\begin{figure}\includegraphics[width=8cm]{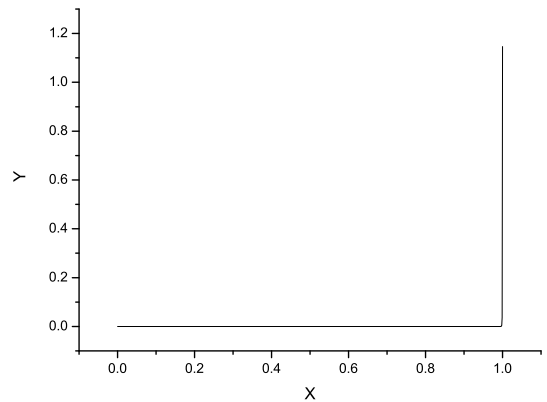}
\caption{The relation between Y$\equiv
10^6\times\overline{\frac{{\rm d}\Sigma}{{\rm d}\Omega_{k'}}}$ and
X$\equiv\theta/\pi$, in the case of $\lambda=$785nm and
$I=10^{19}\mbox{W/m}^2$ for the laser, and energy $E=307$MeV for the
incident electron.}\label{fig2}
\end{figure}

For an unpolarized incident electron beam, if the spin states of the
outgoing particles are not measured, the probability of finding a
photon emitted on direction of $\mbox{\boldmath$k$}'$ is
proportional to the average differential cross section
\begin{equation}\overline{\frac{{\rm d}\Sigma}{{\rm d}\Omega_{k'}}}\equiv
\frac{1}{2}\sum_{i=1,2}\sum_{\sigma=\pm 1,\sigma'=\pm 1}\frac{{\rm
d}\Sigma}{{\rm
d}\Omega_{k'}}(\mbox{\boldmath$e$}'_i).\label{gs70}\end{equation}
Substituting parameters
\begin{equation}\lambda=785{\rm nm},\;\;
I=10^{19}\mbox{W/m}^{2}, \;\mbox{and}\; E=307{\rm
MeV}\label{gs0}\end{equation}into equations (\ref{w2}) and
(\ref{gs70}), we obtain numerical results for the average
differential cross section as a function of $\theta$, which is shown
in the figure \ref{fig2}. The angular distribution for the
$\gamma-$photon emission is very characteristic. Photons concentrate
in the forward direction of the incident electrons, and form a very
sharp peak on this direction. It means an almost mono-directional
emission. The single valued relation (\ref{g5}) between the energy
and the direction of the emitted photon tells us, it also means an
almost monochromatic emission. The stimulated emission makes
positive feedback, and therefore makes a collapse of the
distribution into an almost geometric line with very high intensity.
This is the $\gamma-$ray amplification by stimulated emission of
radiation. Using equations (\ref{g7}) and (\ref{g8}), we find, from
the description at the end of the last section, that the
polarization vector of the forward radiation emitted by electrons is
certainly
\begin{equation}\mbox{\boldmath$e$}'=\frac{1}{\sqrt{2}}
(\mbox{\boldmath$x$}_0-{\rm i}\mbox{\boldmath$y$}_0),\end{equation}
independent of the spin states of the incident electrons. The
radiation emitted by wiggling electrons is therefore completely
polarized. A radiation, even though mono-directional, monochromatic
and completely polarized, does not necessarily mean a laser. A laser
must have a nonzero coherence amplitude, like the amplitude $A$ of
the laser (\ref{16}). It requires that the photon number of the
state is not certain. Theoretically, this is not a problem in our
case. The stimulated emission makes the photon number in the final
electromagnetic state uncertain. The radiation emitted by wiggling
electrons should therefore be a $\gamma-$ray laser, to be confirmed
by experiments.

On the other hand, the coherence amplitude $A=1.5\times10^{-2}mc/e$
in our case here is small.   We have seen, in this limit, the energy
of the photon emitted by wiggling electrons in a laser with ${\cal
N}=1$ approaches the energy of the final state photon in the
corresponding Compton backscattering. We have also seen in huge
number of numerical examples\cite{z4}, that the  average cross
section calculated by eqs.(\ref{w2}) and (\ref{gs70}) approaches the
result of Klein-Nishina formula\cite{kn,n,we} in the same limit. The
results shown in Figures \ref{fig1} and \ref{fig2} are therefore
qualitatively true also for CBS. The radiation generated by the
usual Compton backscattering, in which the incident photons do not
form a coherent state, is also mono-directional, monochromatic and
completely polarized on the forward direction of the incident
electron motion. It means, what one has generated in the CBS
experiments, might already be a $\gamma-$ray laser. It would be
interesting to check this point carefully in old and new
experiments.
\section{Balance between emission and absorption, the intensity of
$\gamma-$ray laser\label{A4} }

Suppose electron-laser collisions happen in a tube of cross section
$S$ and length $L$, with $\sqrt{S}\ll L$. It is the active tube for
the $\gamma-$ray laser generation. Electrons come from an
accelerator, enter the tube at one of its ends, say end 1; then move
to another end, say end 2, and exit from the tube there. The laser
enters the tube at end 2, then propagates along an opposite
direction in the tube to end 1, and exits there. In the tube, beside
the incident laser, there are initial state electrons of density
$n(l)$, the emitted $\gamma-$photons of density $N(l)$, and the
final state electrons of density $n'(l)$, at the point $l$. $l$ is
an one dimensional coordinate of this point, denoting its distance
to the end 1 of the tube. Two competing processes occur in the tube.
One is the emission of the $\gamma-$photon by an initial state
electron, after emission the electron transits to its final state.
Another is the absorption of the emitted $\gamma-$photon by a final
state electron, an initial state electron reappears. The interaction
Hamiltonian $H'_{(ri)}$ in equation (\ref{gs1}) is Hermitian. It
makes the reciprocal theorem valid for the emission and absorption
processes here. In the local approximation we limit $N$ to be the
photon number in an unit volume around the position $l$ under
consideration. It is $N(l)$. In this meaning we have
\begin{eqnarray}&&{\rm d}N=[(N+1)n-Nn']a{\rm d}l/\lambda_c,\label{gs2}\end{eqnarray}
in which $a=\overline{\frac{{\rm d}\Sigma}{{\rm
d}\Omega_{k'}}}/(N+1)$ is a constant independent of position $l$. It
is easily seen, that $n+N$ and $n+n'$ are constants. Taking initial
conditions
\begin{equation}N(0)=n'(0)=0\;\;\;\;\mbox{and}\;\;\;\; n(0)\equiv
n_0,\label{gs3}\end{equation} we have
\begin{equation}N(l)=n'(l)=n_0-n(l).\label{gs4}\end{equation} Substituting them into equation
(\ref{gs2}), we obtain
\begin{equation}\lambda_c\frac{{\rm d}n}{{\rm d}l}=a[2n^2-(3n_0+1)n+n_0^2].\label{gs5}\end{equation}
This is the evolution equation of the process. The solution $n(l)$
under the initial conditions (\ref{gs3}) and, by eq. (\ref{gs4}),
also $N(l)$ and $n'(l)$ are obtained:
\begin{widetext}\begin{equation}n(l)=n_0\frac{\sqrt{n_0^2+6n_0+1}-n_0-1+(\sqrt{n_0^2+6n_0+1}+n_0+1)
\exp(-\sqrt{n_0^2+6n_0+1}\;al/\lambda_c)}{\sqrt{n_0^2+6n_0+1}-n_0+1
+(\sqrt{n_0^2+6n_0+1}+n_0-1)\exp(-\sqrt{n_0^2+6n_0+1}\;al/\lambda_c)}\;,\label{gs6}
\end{equation}\begin{equation}N(l)=n'(l)=2n_0\frac{1-
\exp(-\sqrt{n_0^2+6n_0+1}\;al/\lambda_c)}{\sqrt{n_0^2+6n_0+1}-n_0+1
+(\sqrt{n_0^2+6n_0+1}+n_0-1)\exp(-\sqrt{n_0^2+6n_0+1}\;al/\lambda_c)}\;.\label{gs7}
\end{equation}\end{widetext}

Take parameters (\ref{gs0}), which have been used in making figure
\ref{fig2}, we obtain $\lambda_c/a=337$nm. For tube length $L=1$cm,
$aL/\lambda_c$ is already practically infinity. Eq. (\ref{gs7})
shows, in this case, at the end 2 of the tube, the number density of
$\gamma-$quanta in the generated $\gamma-$ray laser approaches a
constant,  which is about a half of the number density of the
incident electrons. This is a result of the balance between
emissions and absorptions of $\gamma-$quanta by the initial state
and final state electrons respectively in the tube. The intensity of
the output $\gamma-$ray laser is therefore almost determined by the
intensity of the input electron current. For an electron beam of
density $n_0=10^{18}/$m$^3$ and energy $E=307$MeV, which may be
found in many laboratories, the intensity of the output $\gamma-$ray
laser is about $5\times10^{13}$W/m$^2$. It is meaningful.

To generate a more intense $\gamma-$ray laser by the head on
collision between an electron beam and an usual laser, one needs a
more intense electron beam. This situation is some what similar to
the generation of an usual laser by the coherent decay of a dense
aggregate of similarly excited atoms, without pumping. The intensity
of the laser is therefore limited by the limit density of the laser
generating medium. In our case, it is limited by the density of the
electron beam. In the theory presented above, Coulomb interaction
between electrons is totally omitted. This is not permissible if the
density of the electron beam becomes too high. Denote the mean
linear dimension occupied by an electron in the beam by $r$. The
Coulomb force between two neighboring electrons is $\alpha/r^2$ in
the nature units. On the other hand, the electric force of an
applied laser acting on an electron is $eAk$, in which $A$ and $k$
are the amplitude and the circular frequency of the laser
respectively. Therefore, the Coulomb interaction is omissible only
when $r\gg\sqrt{\alpha/eAk}\equiv r_c$. It means the number density
$n$ of the electron beam has to be much less than the critical
density $n_c=r_c^{-3}$. Take parameters (\ref{gs0}), we obtain
$n_c\approx 10^{29}/$m$^3$. To be save, we take $n_0=10^{23}/$m$^3$.
It may generate a $\gamma-$ray laser of intensity $10^{18}$W/m$^2$,
if technical difficulties may be overcome. For too dense electron
beams, the space charge effect becomes important. To consider the
collision between a laser and a dense electron beam with serious
space charge, we have to essentially modify the theory presented
above.
\section{Multi-section lasers, pumping and resonator, linear
and cyclic intensifiers\label{A5}}

\begin{figure}\includegraphics[width=6cm]{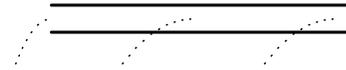}
\caption{Sketch map for the multi-section active tube. The space
between two parallel black lines shows the inside of the tube, dote
lines show the trajectories of the electron injection. }\label{fig3}
\end{figure}

The space charge problem may also be solved in a technical way. It
is to divide one high density electron beam into many sub-beams with
much lower densities, and inject them into the active tube from
different entrances at different times. Entrances are opened on the
wall of the tube, distributed along its length direction. The
electron sub-beams are injected through them by suitably designed
magnets so that the injected electrons move along central axis of
the tube. The injection time is also controlled to let the injected
electrons meet the already emitted $\gamma-$quanta as soon as they
arrive the central axis, so that the stimulated emissions may
continue. Figure \ref{fig3} is a sketch map of the designation. We
also open exits for electrons, one by one with the entrances, on the
wall of the tube. After participating the emission and absorption of
$\gamma-$quanta, electrons are removed from these exits by suitably
designed magnets, just before new electrons being injected through
the corresponding entrances.

The active tube is now constituted by its sections, each begins from
an entrance and ends at an exit. The tube analyzed above is a single
section one.  We may analyze each section in a multi-section tube in
a similar way. Let us concentrate on one section of the tube. $l$
denotes the distance between a point in the section and the entrance
of this section. Start from equation (\ref{gs2}). We need a
modification for the initial conditions (\ref{gs3}). The
$\gamma-$quantum number density at the beginning of a section is
$N(0)\equiv N_0\geq 0$ in general, the equal mark works only for the
beginning section. Equation (\ref{gs4}) now is substituted by
$N(l)=N_0+n'(l)=N_0+n_0-n(l)$. The evolution equation is generalized
to be\begin{equation}\lambda_c\frac{{\rm d}n}{{\rm
d}l}=a[2n^2-(2N_0+3n_0+1)n+n_0(n_0+N_0)].\label{gs8}\end{equation}
This is slightly more complicated than equation (\ref{gs5}), but is
still analytically solvable.  An analysis similar to that made in
the section \ref{A4} shows, for the case with parameters
(\ref{gs0}), in a section of length $L\!\sim$ centimeters, after a
balance between emissions and absorptions, the net increase of
average photon number density of the $\gamma-$ray laser at the end
of the section,  is still a noticeable fraction of the number
density $n_0$ of newly injected electrons.

At the beginning of the new section, the initial state electron
number density restores its original value $n_0$, and the final
state electrons are removed. It is similar to the pumping in the
usual laser generation. The intensity of the laser is therefore
multiplied by this mechanism. A $\kappa$ section tube makes $\kappa$
times pumping, and therefore intensifies the $\gamma-$ray laser,
roughly say, by a factor  $\kappa$. For the applied laser with
parameters (\ref{gs0}), the section length is of the order of
centimeters, and the length of a 100 section tube is of the order of
meters. It is acceptable. For the incident electron beam of density
$n_0=10^{18}/$m$^3$, the intensity of the final output $\gamma-$ray
laser is of the order of $10^{15}$W/m$^2$. If the density of the
incident electron beam becomes $n_0=10^{23}/$m$^3$, we will have a
$\gamma-$ray laser of intensity $10^{20}$W/m$^2$.

The intensity of a multi-section $\gamma-$ray laser is also
influenced by the intensity of the applied laser. More intense
incident laser gives larger emission probability for an electron
wiggling in it, which in turn gives larger coefficient $a$ in the
evolution equation (\ref{gs8}) and shorter section in the active
tube. An active tube of given length may therefore contain more
sections and give higher intensity for the $\gamma-$ray laser.

A long (several meters or several ten meters), thin and well
collimated many section active tube also plays a role here, which
was played by a resonator in the usual laser generation. It selects
a definite mode of $\gamma-$radiation, with definite direction,
therefore definite wavelength and definite polarization, to amplify
by stimulated emissions. Other modes are washed out from the
amplification,  because of the positive feedback. A longer and
thinner active tube makes the quality of the $\gamma-$ray laser
better. We call this structure a linear intensifier of the
$\gamma-$radiation.
\begin{figure}\includegraphics[width=6cm]{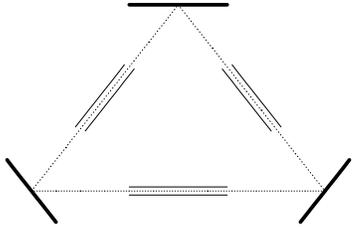}
\caption{Sketch map for the cyclic intensifier. The spaces between
two parallel thin gray lines show the inside of the active tube,
dote lines show the paths of the emitted photons, thick black lines
show crystals making Bragg reflection of the emitted $\gamma-$ray.
}\label{fig4}
\end{figure}

One may change the direction of a soft $\gamma-$radiation by
Bragg-reflection, and therefore may arrange the sections of the
active tube for amplifying this radiation in a cyclic way. In
principle, this structure may make a huge number of amplifications
for the $\gamma-$ray by stimulated emissions in a finite space, and
reach a very high intensity, if the reflection efficiency is high
enough. We call this structure a cyclic intensifier of the
$\gamma-$radiation. A sketch map of it is shown in figure
\ref{fig4}. Unfortunately, the Bragg-reflection works only for soft
$\gamma-$radiations with wavelengths of the order of the lattice
constants of some crystals, it is of the order of Angstroms. Only
lasers composed of soft $\gamma-$photons may be intensified in this
way. For an example, in the head-on collision between an usual laser
and a beam of electrons with
parameters\begin{equation}\lambda=785{\rm nm},\;\;
I=10^{19}\mbox{W/m}^{2}, \;\mbox{and}\; E=7.68{\rm
MeV},\label{gss}\end{equation}the energy of the output photon on the
forward direction of incident electrons is 1.424KeV, it is in the
soft $\gamma-$ray range. We expect the $\gamma-$ray of this kind may
be intensified by the cyclic intensifier to a rather high intensity.

\section{Confirmation of the coherence\label{A6}}

Finally we have to confirm that the generated and amplified
$\gamma-$ray is indeed a laser. The most direct way for this purpose
is to measure its coherence amplitude $A$, and to see whether it is
nonzero or not. Equation (\ref{g5}) of ${\cal N}=1$ and equation
(\ref{g55}) help us to do so. Taking reciprocals of two sides in
these equations, subtracting the result of the latter from that of
the former, we obtain
\begin{equation}\frac{\Delta\lambda'}{\lambda'_0}\!\equiv\!\frac{\lambda'\!-\!\lambda_0'}{\lambda_0'}\!=\!
\frac{\left(eA\sin\frac{\theta}{2}\right)^2}{(E-p_z)[E+k-(p_z+k)\cos\theta]}.\label{g60}
\end{equation} $\lambda'\equiv 2\pi/k'$ and $\lambda'_0\equiv 2\pi/k_0'$ are wavelengths of
outgoing radiations in the electron-laser collision and in the usual
Compton backscattering respectively. The coherence amplitude $A$
simply manifests itself in the wavelength shift
$\Delta\lambda'/\lambda_0'$ of the outgoing radiations.

Suppose the output soft $\gamma-$radiation generated in the
electron-laser head on collision with parameters (\ref{gss}) is
intensified to a high intensity of $I=10^{26}$W/m$^2$ by a cyclic
intensifier. A beam of electrons of energy $E=5.135$MeV is injected
into the radiation along its propagation direction. Electrons
interact with the radiation and emit ultraviolet radiation on the
direction against the $\gamma-$ray propagation. Equations
(\ref{g9}), (\ref{g5}) and (\ref{g60}) show that, if the
$\gamma-$radiation is completely coherent, the wavelength of the
emitted ultraviolet radiation would be $\lambda'=351$nm, and the
wavelength shift due to the nonzero coherence amplitude $A$ would be
$2.77\times10^{-3}$. They may be detected and checked by
experiments, using normal optic techniques, well developed about a
century ago, for the fine structure research in the atomic
spectroscopy.

It is possible, that the $\gamma-$ray is constituted by parts. One
forms a coherent state, it is the laser. Another is a non-coherent
aggregate of photons. The variable $I$ in the equation (\ref{g9}) is
the intensity of the coherent part. To be definite, we call it the
coherent intensity. Combining equations (\ref{g9}) and (\ref{g60}),
we see
\begin{equation}\frac{\Delta\lambda'}{\lambda'_0}=\frac{\alpha\lambda^2\sin^2\frac{\theta}{2}}
{\pi(E-p_z)[E+k-(p_z+k)\cos\theta]}\,I \label{g61}\end{equation} in
the nature unit system. The wavelength shift is proportional to the
coherent intensity of the radiation. In the above example, we
assumed that the radiation intensity equals its coherent intensity.
The radiation, as a whole, is a laser. If there are non-coherent
photons, the coherent intensity will be lower than the total
intensity. It results in a smaller wavelength shift. In other words,
smaller wavelength shift means incomplete coherence. If in the above
example a smaller wavelength shift, say $\Delta
\lambda'/\lambda'_0=2.77\times10^{-4}$ is observed, one has to
conclude that only one tenth emitted photons form a coherent state.
Other photons are non-coherent.
\section{Conclusions\label{A7}}

Based on quantum electrodynamics, we have shown that, both in the
head on electron-laser collision and in the Compton backscattering,
the emitted photons may form a $\gamma-$ray laser. The competition
and balance between the emission and reabsorption of $\gamma-$quanta
by electrons determine the evolution and therefore the output
intensity of the $\gamma-$ray laser.

In electron-laser collisions, incident electrons, wiggled by the
applied laser, emit $\gamma-$photons. The stimulated emissions
amplify the $\gamma-$radiation, form the $\gamma-$ray laser. In this
sense, the generated $\gamma-$ray laser is a Free Electron Laser
(FEL). The new here is that the electron motion in the background
laser is governed and exactly solved by quantum mechanics.
Therefore, it is a Quantized Free Electron Laser (QFEL). This
mechanism may be understood in an usual way as well. The incident
electrons are higher state systems in the background laser. They
transit into final (lower) state and emit $\gamma-$photons. The
stimulated emissions again make the $\gamma-$ray laser. Two
understandings of the $\gamma-$ray laser generation are unified in
quantum theory. In this way, ideas in the usual laser theory, such
as pumping, resonator, positive feedback, and so on are usable in
the analysis. By use of them, we propose some ways to intensify and
qualify $\gamma-$ray lasers. Analytical and numerical estimations
show that the possible generated $\gamma-$ray laser may be intense
enough to be detected and  researched. A way for measuring the
coherence amplitude $A$, and therefore confirming the coherence of
the $\gamma-$ray laser is proposed.

Of course, it is not easy to work out these proposals. Many
technical problems have to be solved before their realization. For
an example, the theory is developed under an ideal condition, in
which incident electrons have identical energies and move on
identical directions, and the background laser is a plane wave. In
reality, one can infinitely approach this condition, but cannot
reach it. The experience of the usual laser generation tells us,
there is a critical bound around the ideal condition. Crossing over
the bound makes the positive feedback of the stimulated emission
become dominating, and the emitted photons collapse into a laser.
This is a phase transition. To find the critical bound and realize
the phase transition for a $\gamma-$ray laser would be a heavy work,
mainly experimental.\vskip 0.1cm

\noindent $Acknowledgment$ The work is supported by the Nature
Science Foundation of China with grant number 10875003.

\end{document}